\documentclass[onecollarge,natbib]{svjour2}
\bibpunct{[}{]}{;}{n}{}{,} 
\smartqed  
\usepackage{graphicx}
\usepackage{wrapfig}
\usepackage{sidecap}

\journalname{Few-Body Systems}

\mathchardef\bbkappa="7114
\mathchardef\bbrho="711A
\mathchardef\bbsigma="711B
\mathchardef\bbtau="711C
\mathchardef\bbvarrho="7125
\mathchardef\bbvarsigma="7126
\mathchardef\bbxi="7118

\def\be{\begin{eqnarray} &&}

\def\nonu{\nonumber \\ &&}
\def\ee{\end{eqnarray}}
\newcommand\la{\langle}
\newcommand\ra{\rangle}

\def\beq{\begin{equation}}
\def\eeq{\end{equation}}
\usepackage{amsbsy,amsmath}

\def \brho {{\mbox{\boldmath$\rho$}}}
\newcommand {\br} {{\bf r}}

\newcommand{\bq}{\begin{eqnarray}}
\newcommand{\eq}{\end{eqnarray}}

\newcommand {\vece}[1]{\overset{_\rightarrow}{#1}}
\begin{document}
\title{Polarized $^3$He target and final state interactions in SiDIS}
%
%

\author{Alessio Del Dotto \and
        Leonid Kaptari \and
Emanuele Pace \and
	    Giovanni Salm\`e \and
       Sergio Scopetta
}

\institute{
A. Del Dotto \at
   INFN Sezione di Roma, Italy \\
   Thomas Jefferson National Accelerator Facility,
   Newport News, VA 23606, USA\\
\email{alessio.deldotto@iss.infn.it}
\and
L. Kaptari \at
Bogoliubov Lab. Theor. Phys., 141980, JINR, Dubna, Russia
\and
E. Pace \at
 Dipartimento di Fisica, Universit\`a di Roma ``Tor Vergata'' and INFN, Roma 2, Italy
\and
G. Salm\`e \at
           INFN Sezione di Roma, Italy
\and
S. Scopetta \at
 Dipartimento di Fisica e Geologia, Universit\`a di Perugia and INFN, 
Sezione di Perugia, Italy 
          }
\date{Received: date / Accepted: date}

\maketitle

\begin{abstract}
 {Jefferson Lab (JLab) is starting a wide experimental program aimed at  studying the
neutron's structure, with a great emphasis on the extraction of   the parton
transverse-momentum distributions (TMDs). To this end, 
 Semi-inclusive deep-inelastic scattering
(SiDIS) experiments on polarized $^3$He will be carried out, providing,} together
with proton and deuteron data, a sound flavor
decomposition of the TMDs.
Given the expected high statistical accuracy, it is crucial
to disentangle nuclear and partonic degrees of freedom
to get an accurate theoretical description of both initial
and final states.
{In this contribution, a preliminary  study of the Final
State Interaction (FSI) in the standard SiDIS, where a
pion (or a Kaon) is detected in the final state is presented, in view of constructing a realistic
description of the nuclear initial and final states.}
\end{abstract}

\section{Introduction}
\label{intro}

In order to extend our knowledge of the nucleon structure, we need to access the three-dimensional 
picture of neutron and proton in terms of the degrees of freedom of their constituents. 
Information on the three-dimensional momentum space of the nucleon can be obtained extracting the 
quark transverse momentum distributions (TMDs) \cite{uno} from the so-called single spin asymmetries 
(SSAs), which can be measured in polarized semi-inclusive deep inelastic scattering (SiDIS). {Indeed, by exploiting SiDIS 
off transversely polarized target the Sivers and Collins contributions can be selected \cite{uno}, showing from the present data on 
 ${\vece p}(e,e'\pi)x$ \cite{due} and $\vece D(e,e'\pi)x$ \cite{tre}  a 
strong flavor dependence of TMDs. To extend such a study, measurements with a $^3\vece{\mathrm{He}}$ target become compelling (see Ref.  \cite{tre1},
for the first data at 6 GeV)
 and, in 
the close future,   highly accurate  experiments  are 
planned at {12-GeV Jlab} }\cite{quattro}. 
The neutron data will be extremely important to achieve the flavor separation of the TMDs \cite{cinque}; 
therefore to obtain a reliable information one has to take into account the nuclear structure of $^3\mathrm{He}$ 
considering also the final state interaction (FSI) between the {detected} pion and the remnant debris.
In the following we report our {preliminary results on this issue}.    

\vspace{-10pt}

\section{The polarized $^3$He nucleus  as an effective neutron target}
\label{sec:1}

A polarized $^3$He nucleus  is an ideal target to study the neutron, since at 
a 90\% level it is equivalent to a polarized neutron. For disentangling the nucleon
structure from the  
dynamical nuclear effects, one can 
adopt  an approach {primarily} based on the spin-dependent spectral function of 
$^3$He, ${\rm P}_{\sigma,\sigma^\prime} (\vec p, E)$, (see, e.g.  \cite{Ciofi1})
that yields
the probability distribution to find a nucleon with given missing energy {(the spectator pair is interacting)},
three-momentum and polarization inside the nucleus.
By using  this formalism, {known as Plane Wave Impulse Approximation (PWIA),} one can safely extract \cite{Ciofi2} the 
neutron longitudinal asymmetry, $A_n $,
from the corresponding $^3$He  observable, $A_3^{exp}$, 
obtained from   the reaction
 $^3\vece{{\rm He}}( \vece{e},e')X$ in DIS regime, i.e.   
\be
  {{A_n }}\simeq  \left 
( {A^{exp}_3} - 2 
{p_p} f_p
{{A^{exp}_p}} \right )/(p_n f_n )\label{dis}\ee
with  $p_{n(p)}$  the
neutron (proton) effective polarization inside the polarized $^3$He, and
$f_{n(p)}$,  the dilution factor. Realistic values
of $p_n$ and $p_p$ are  
$ {{p_p}} = -0.023  $, 
${{p_n}}= 0.878 $  (see, e.g., \cite{Ciofi2,Sco}).
  In \cite{Sco}, an analogous extraction was applied to the SSA of a
  transversely polarized $^3\vece{{\rm He}}$ target, obtained 
  from
the  process $^3\vece{{\rm He}}( e,e'\pi)X$, in order to obtain the 
SSA of a transversely polarized neutron. 
 In PWIA and adopting  the Bjorken limit, the SSAs of  $^3\vece{\rm { He}}$
are a  convolution of  
$ {\rm P}_{\sigma,\sigma^\prime} (\vec p, E)$, and the nucleon SSAs, that in turn
are convolutions of suitable 
  TMDs
 and  fragmentation functions (FF), phenomenologically describing  the hadronization
 of the hit quark. The same extraction procedure has been also applied in combination with a Monte Carlo 
simulating the kinematics of the experiment E12-09-018 \cite{sei}.
The main ingredients for the extraction were: i) a realistic ${\rm P}_{\sigma,\sigma^\prime} (\vec p, E)$ for the $^3 \mathrm{He}$, 
obtained using the AV18 interaction; 
ii) parametrizations of data for TMDs and FFs, whenever available; {iii)} models for the unknown TMDs and
FFs. {Within this framework}, the  extraction formula (\ref{dis}) works well 
even in the case of SiDIS, at least at the present statistical accuracy of the existing data. Indeed
Eq. (\ref{dis}) has been used by the 
JLab Hall A Collaboration to extract, for the first time, the Collins and Sivers moments from a transversely polarized $^3\mathrm{He}$ target \cite{tre1}.
It is important to point out that the existing measurements are limited in statistical accuracy and kinematics coverage, therefore 
an extensive program of high precision measurements of SiDIS off $^3\mathrm{He}$ will be part of the JLab program at 12 GeV \cite{quattro}. 
The expected statistical accuracy is of the order of percent in a wide range of multi-dimensional kinematical binning; for this reason the 
PWIA could be no longer sufficient and the FSI, {(not considered in PWIA)} may have a non-negligible role. Our aim is to add this effect {to the formalism}, obtaining a 
distorted spin-dependent spectral function of the $^3\mathrm{He}$.   
\vspace{-10pt}
\section{Beyond PWIA: the generalized eikonal approximation}
\label{sec:2}
\begin{wrapfigure}{l}{0.3\textwidth} 
\vspace{-20pt}
\includegraphics[width=5cm]{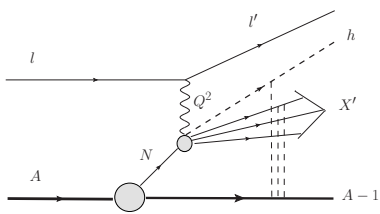}
 \caption{Interaction between the $(A-1)$ spectator system {(fully interacting)} and the debris
produced by the
absorption of a virtual photon by a nucleon in the nucleus.}
\vspace{-20pt}
\label{fig:1}
\end{wrapfigure}
The JLab SiDIS experiments will exploit an electron beam at 8.8 and 11 GeV off $^3\mathrm{He}$ polarized gaseous target; the relative energy between the $(A-1)$ system and the system of the detected pion and the remnant (see Fig. \ref{fig:1}) is a few GeV therefore the FSI can be treated within a generalized eikonal approximation framework (GEA). The GEA was already successfully applied to unpolarized SiDIS \cite{sette}, and in a recent paper the {\it distorted} spin-dependent spectral function has been calculated for the {\it spectator} SiDIS, where a slow $(A-1)$ system nucleon system, acting as a spectator of the photon-nucleon interaction, is detected, while the produced fast hadron is not \cite{otto}.
In the following we are going to report preliminary results for the usual SiDIS, where all the possible state of the two-nucleon spectator system have to be considered. 
\vspace{20pt}
The {\it distorted} spin-dependent spectral function for a polarized $^3\mathrm{He}$ target can be written as 
{\be
S^{N \,{\bf S}_3}_{\lambda\lambda'}(E,{\bf p}_{mis})=
\nonu =
\sum_{f_2} 
\sum \! \!\! \!\! \!\! \!\int_{~\epsilon^*_2}\rho\left(
\epsilon^*_2\right)\,
{ \tilde {\cal O}}_{\lambda\lambda'}^{N \, {\bf S}_3 \, f_2}
(~\epsilon^*_2,{\bf p}_{mis})
\,
\delta\left( { E+ M_3-m_N-M^*_2-T_2}\right)
\ee
 with
 the  product of distorted overlaps defined by
\be
{ \tilde {\cal O}}_{\lambda\lambda'}^{N \, {\bf S}_3 \, f_2}
(\epsilon^*_2,{\bf p}_{mis})=
\nonu =
\la \lambda,
\phi_{\epsilon_2^*}^{f_2}(\br)
e^{-i{\bf p}_{mis}\brho} {\cal G}(\br,\brho)
|
\Psi_3^{{\bf S}_3}(\br,\brho)\ra
\la \Psi_3^{{\bf S}_3}(\br',\brho')| \lambda', {\cal G}(\br',\brho')
\phi_{\epsilon_2^*}
^{f_2}
(\br')
e^{-i{\bf p}_{mis}\brho'}\ra.
\label{overfsi}
\ee }   
{where i) $\rho\left(\epsilon^*_{2}\right)$ is the density of the spectator pair  with intrinsic 
energy $\epsilon^*_{2}$, ii) $|\Psi_3^{{\bf S}_3}(\br,\brho)\ra$ is the ground state of the  3-nucleon system
 with polarization ${\bf S}_3$} iii) $E$ is the usual missing energy 
$E = \epsilon^*_{2} + B_3$, if the kinetic energy $T_2$ of the spectator pair is disregarded and ${\bf p}_{mis}$ is the three momentum of the spectator pair. 

\begin{SCfigure}
  \centering
  \caption{The $^3$He spectral function, for the neutron,
in the unpolarized case, as a function of $p_{mis}=|{\bf p}_{mis}|$ and
of the removal energy $E$, in PWIA (full lines) and with
FSI taken into account within GEA framework (dotted lines).
The spectral functions are shown for the
values of $E$ and $p_{mis}$ that contribute to the calculus 
of the light-cone momentum distribution $\mathrm{f^n(\alpha)}$ for $\alpha \equiv A(p_N\cdot q)/(P_3\cdot q) = 0.86$ 
in the SiDIS cross section at ${\cal E}=$ 11 GeV and $Q^2 \simeq 7.6$ (GeV/c)$^2$; (see 
also Ref. \cite{dieci} for more details).}
\includegraphics[width=0.45\textwidth]{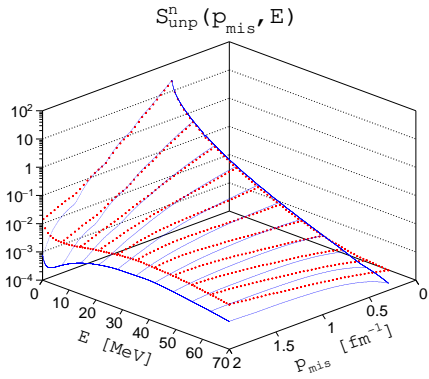}
\label{fig:2}
\end{SCfigure}

The Glauber operator in Cartesian coordinates is given by 
\be
{\cal G}(\br_1,\br_2,\br_3)={\cal G}(\br,\brho)=\prod\limits_{i=2,3}
\left[ 1-
\theta(\br_{i||}-\br_{1||})
\Gamma \left( \br_{i\perp}-\br_{1\perp},\br_{i||}-\br_{1||}
\right) \right ]~,
\label{gl}
\ee
where $\hat\br_{i\perp}$ and  $\hat\br_{i||}$ are the perpendicular and the parallel components of $\br_i$ with respect to 
the direction of the debris. The profile function 
\be
\Gamma({\br_{i\perp}},\br_{i||})\, =\,\frac{(1-i\,\eta)\,\,
\sigma_{eff}(\br_{i||})} {4\,\pi\,b_0^2}\,\exp \left[-\frac{{\br^2_{i\perp}}}{2\,b_0^2}\right]~~~,
 \label{eikonal}
\ee
unlike the standard Glauber approach depends not only on the impact parameter but also on the longitudinal 
separation trough an effective cross section. This expression has been already used in our previous works 
{(for details on the model  see  Ref. \cite{otto,nove})}.
To recover the PWIA formulation one has simply to put ${\cal G}\equiv 1$. In Fig. \ref{fig:2} a {preliminary} plot of the 
$^3\mathrm{He}$ distorted and undistorted spectral function, for the neutron, in the unpolarized case is shown.  

\vspace{-10pt}
\section{Good news on the extraction of the neutron's SSAs}
\label{sec:3}
\begin{figure}[h]
\centering
\includegraphics[scale=0.34]{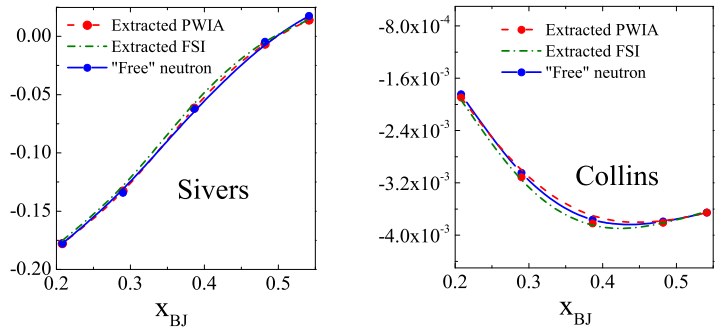}
\caption{Neutron asymmetries extracted trough Eq. (1) from the Sivers (Left panel) and Collins (Right panel) asymmetries with and without FSI, in actual kinematics of JLab at 12 GeV \cite{quattro}. Preliminary results to appear in \cite{dieci}.}
\label{fig:3}
\end{figure}
The relevant part in the extraction of the transversely polarized neutron's SSAs is the transverse spectral function, given by
\be
{\cal S}^{N \, \perp}(E,{\bf p}_{mis})={\cal S}^{N \, \frac12 -\frac12}_{\frac12 -\frac12}(E,{\bf p}_{mis})
+ {\cal S}^{N \, -\frac12 \frac12}_{\frac12 -\frac12}(E,{\bf p}_{mis})
\label{trspectr}
\ee

 In general ${\cal S}^{\perp (PWIA)}$ and ${\cal S}^{\perp (FSI)}$ can be quite different {and the effective polarizations are respectively
  $p^{PWIA}_p = -0.023$ and $p_n^{PWIA} = 0.878$; $p^{FSI}_p = -0.026$ and $p_n^{FSI} = 0.760$.} Then $p_{p(n)}$ with and without 
  FSI differ by 10-15\%. Nevertheless, in Eq. (\ref{dis}) the effective polarizations occur in products with the dilution factor and to a large 
  extent $p_p^{PWIA}f_p^{PWIA}\approx p_p^{FSI}f_p^{FSI}$ ,  $p_n^{PWIA}f_n^{PWIA}\approx p_n^{FSI}f_n^{FSI}$ \cite{undici}. {Such a fortunate case allows
  one to safely adopt the usual extraction, as shown {by the preliminary results} in Fig. \ref{fig:3}, and therefore the goal of a sound flavor
  decomposition of TMDs seems quite affordable}.    

As a next step, we plan to include the FSI in our Light-Front relativistic description of the $^3\mathrm{He}$, already employed to evaluate the
 relativistic effects in SiDIS processes in PWIA (see \cite{dodici}).  

%
%
%

\end{document}